\documentstyle[floats,epsf,prb,aps,twocolumn]{revtex} 
\begin{document}

\title{Skipping orbits and enhanced resistivity \\in large-diameter 
InAs/GaSb antidot lattices} 
 
\author{J.\ Eroms\cite{adresse}, M.\ Zitzlsperger, D.\ Weiss} 
\address{Universit\"at Regensburg, D-93040 Regensburg, Germany,} 
 
\author{J.\,H.\ Smet, C.\ Albrecht} 
\address{Max-Planck-Institut f\"ur Festk\"orperforschung, 
         Heisenbergstra\ss e 1, D-70569 Stuttgart, Germany} 
 
\author{R.\ Fleischmann} 
\address{Max-Planck-Institut f\"ur Str\"omungsforschung, 
         Bunsenstra\ss e 10, D-37073 G\"ottingen, Germany} 
 
\author{M.\ Behet, J.\ De Boeck, G.\ Borghs} 
\address{IMEC, Kapeldreef 75, B-3001 Leuven, Belgium} 
 
\date{\today} 

\author{\begin{minipage}[t]{15cm}\small We investigated the 
magnetotransport properties of high mobility InAs/GaSb antidot 
lattices. In addition to the usual commensurability features at low 
magnetic fields we found a broad maximum of classical origin around 
$2.5\,$T. The latter can be ascribed to a class of rosetta type 
orbits encircling a single antidot. This is shown by both a simple 
transport calculation based on a `classical' Kubo formula and an 
analysis of the Poincar\'e surface of section at different magnetic 
field values. At low temperatures we observe weak $1/B$-periodic 
oscillations superimposed on the classical maximum. 
\end{minipage}} 

\maketitle 

\label{intro}

Periodically modulated two-dimensional electron gases (2DEGs)
offer the possibility to study electron motion in artificially 
tailored periodic potentials.\cite{andospringer} 
If the modulation potential is strong enough to deplete parts of the 
2DEG around the potential maxima the system is called an antidot 
(AD) array. 
Antidot lattices are model systems to study both the classical 
chaotic motion 
of 
electrons in the potential landscape as well as resulting 
bandstructure effects.\cite{andospringer} Investigations so far have 
essentially focused on the low field regime where the low-temperature 
magnetoresistance reflects peaks 
whenever the classical cyclotron diameter $2R_c$ fits around a 
(geometry-dependent) specific number of antidots.\cite{Antidots} 
The last maximum 
appears in the resistivity when $2R_c$ is equal to the period $a$ 
of the array. 
Here we focus on novel structure appearing at slightly higher fields 
corresponding to $B$-values where $2R_c < a$ holds. While antidot 
lattices are 
usually based on GaAs/AlGaAs heterojunctions, we used here the 
material system InAs/GaSb to fabricate antidot arrays. Due to the 
pinning of the Fermi energy within the conduction band at 
open surfaces of InAs,\cite{Leitungsband} depletion regions around 
the antidots should be 
significantly smaller than in GaAs based systems. This should allow, 
due to the 
resulting steep potential posts, to fabricate very short period 
antidot arrays 
or, as it is the case here, the fabrication of large antidots with 
small constrictions between neighboring antidots. 
 
\begin{figure}[htb] 
\begin{minipage}{12cm} 
\epsfxsize=8.6cm 
\epsfbox{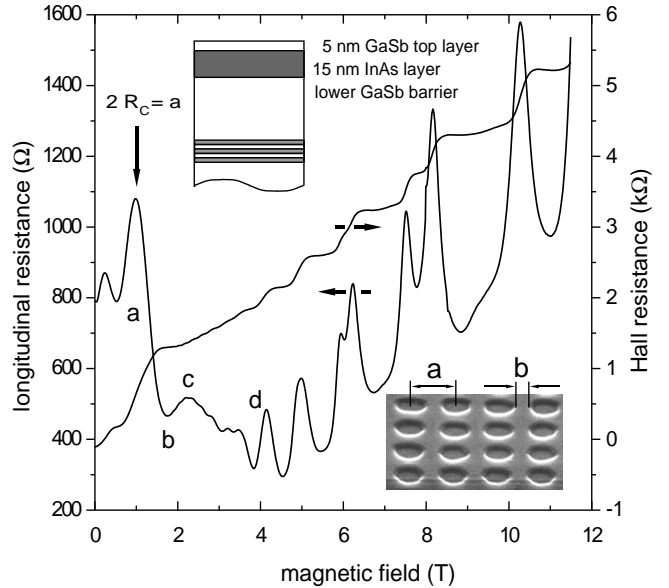} 
\end{minipage} 
\vspace{0.2cm} 
\caption{ Magnetoresistance and Hall resistance of an InAs/GaSb 
antidot lattice. 
Lower inset: A SEM picture of the sample. 
Upper inset: Layer sequence of the heterostructure as described in 
the text. 
Letters {\sf a} to {\sf d} denote $B$-field positions of interest, 
which are analysed numerically in Fig.~\ref{Poincare}. 
} 
\label{uebersicht} 
\end{figure}

Our samples are fabricated from InAs/GaSb heterostructures grown on 
undoped 
GaAs substrates by molecular beam epitaxy. The 
epitaxial layers on the substrate consist of a $1.1\,\mu $m AlGaSb 
layer, a 
$0.5\,\mu $m thick GaSb buffer followed by a ten period 
AlSb/GaSb superlattice, the GaSb barrier of $50 \,\rm{nm}$ thickness, 
the  $15 \,\rm{nm}$ InAs quantum well and a $5 \,\rm{nm}$ GaSb top 
layer (see upper inset in Fig. \ref{uebersicht}). 
While the heterostructure is nominally undoped, the carrier 
concentration $n_s$ can be adjusted by the top layer 
thickness.\cite{Altarelli,Nguyen1,Nguyen2} Typical values of $n_s$ 
in our unpatterned samples were about 
$1.5 \cdot 10^{12} \,\rm{ cm}^{-2}$. Hall-bars 
were defined by standard photolithography and wet chemical etching. 
\cite{Suppe} It turned out that the 
alkaline photoresist developer attacks the GaSb-layer and decreases 
the carrier mobility from about $400\; 000\, \rm{cm^2/Vs}$ to about 
$100\; 000\, \rm{cm^2/Vs}$. 
An annealing step at ${270^{\rm{o}}}$C  in a continuous flow of 
forming gas restored the 
mobility to about $300\; 000\, \rm{cm^2/Vs}$. This corresponds to a 
mean free 
path of $6\,\mu$m which is much larger than the lattice period 
$a=380\,$nm, thus making commensurability features observable. 
Since InAs does not form a Schottky barrier to metals, it was not 
necessary to use alloyed contacts. Instead of gold, which proved to 
react with GaSb in the final annealing step, we used evaporated 
indium contacts. The antidots were defined with electron beam 
lithography and etched with the 
solution described above. A scanning electron micrograph of the 
samples 
showed the antidots 
to be uniform in size and shape with an AD diameter of 
$250\,\rm{nm}$. 
 
The four-point magnetotransport measurements were carried out 
at temperatures ranging from 
$1.5\,$K to $40\,$K using standard lock-in techniques. Utilising the 
negative persistent photo-conductivity of InAs quantum 
wells,\cite{NPPC} the carrier 
density could be varied by illuminating the sample with a red LED. 
The magnetoresistance and Hall resistance traces at a carrier 
density of $1.33 \cdot 10^{12} \,\rm{ cm}^{-2}$ and a temperature of 
$1.5\,$K 
are shown in Fig.~\ref{uebersicht}. 
At low 
fields two commensurability peaks in the magnetoresistance curve can 
be observed. The peak at about 
$1\,$T belongs to trajectories going around one antidot, the other
 one at $0.25\,$T corresponds to a deformed cyclotron orbit around 
four antidots. In the simplest model, the enhanced resistivity at
these field values can be ascribed to electrons pinned on 
commensurate cyclotron orbits, which cannot carry 
current.\cite{Antidots,Ragnar}
At fields above $3.5\,$T strong Shubnikov-de Haas (SdH) oscillations 
appear and from $6\,$T spin-splitting of the Landau levels is 
resolved. 
By comparison to an unpatterned reference sample, 
 we can conclude that 
the quality of the 2DEG was not seriously affected by the patterning 
 process. 
In the region between $1.5\,$T and $3.5\,$T we find 
a new broad peak structure whose origin will be discussed below. 
In earlier work, a corresponding shoulder was ascribed to 
etch-induced defects.\cite{toebben} 
For a sample with triangular antidots, a high-field shoulder at 
$2R_c = a/2$ was explained by electrons being reflected on the 
straight antidot edges.\cite{Schulter}

\begin{figure}[!htb] 
\begin{minipage}{12cm} 
\epsfxsize=8.6cm 
\epsfbox{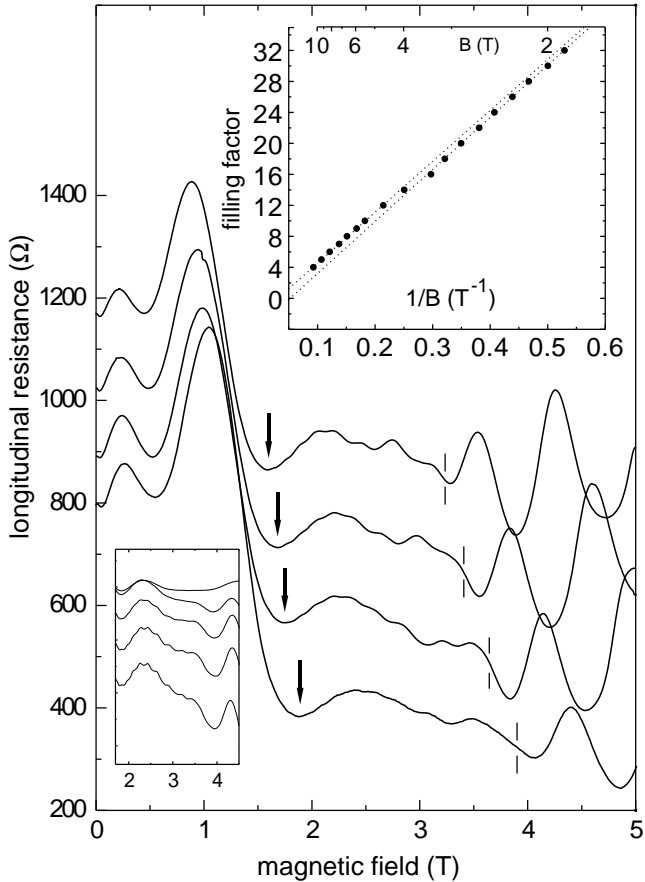} 
\end{minipage} 
\vspace{0.2cm} 
\caption{ 
Density dependence of the magnetoresistance. Electron densities 
(from top to bottom): $1.13 \cdot 10^{12} \,\rm{ cm}^{-2}$, 
$1.24 \cdot 10^{12} \,\rm{ cm}^{-2}$, 
$1.33 \cdot 10^{12} \,\rm{ cm}^{-2}$, 
and $1.42 \cdot 10^{12} \,\rm{ cm}^{-2}$. 
Arrows indicate the beginning of the new structure, vertical lines 
mark the onset of strong SdH oscillations. Lower inset: Temperature 
dependence of the features at 
$n_s=1.42 \cdot 10^{12} \,\rm{ cm}^{-2}$. 
 Temperatures are $1.5\,$K, $3\,$K, $10\,$K, $20\,$K, and $40\,$K 
from bottom to top. Graphs are offset by $ 100 \, \Omega$ for 
clarity. 
Upper inset: $1/B$-position of the minima of the quantum 
oscillations. 
Filling factors $\nu $ correspond to the carrier density determined 
at high fields. 
A magnetic field axis is included for guidance.
} 
\label{dichte} 
\end{figure}

Figure~\ref{dichte} shows the dependence of the magnetoresistance 
traces on the carrier density (changed by illumination) and 
temperature (inset). The novel peak structure shifts similar to the 
main commensurability maximum at $2R_c = a$. This suggests, together 
with the weak temperature dependence of the peaks (see lower inset 
of Fig.~\ref{dichte}), a classical origin of the phenomenon. 
Associating the magnetic field value at which the broad maxima start 
to appear (marked by arrows) with the matching condition $R_c = b$  
we find a value of $b = 110$\,nm. This value $b$ corresponds to the 
distance between neighboring antidots in our device. This finding is 
consistent with the assumption that the broad maximum is associated 
with rosetta shaped orbits skipping around one antidot, as we will 
show below. 
 
Superimposed on the classical peak we find $1/B$-periodic quantum
oscillations. While the periodicity of these oscillations is similar 
to the period of the SdH-oscillations at higher magnetic field, 
there is a distinct transition between the high-field oscillations 
and the ones superimposed upon the classical maximum. This can be 
seen in the upper inset of Fig.~\ref{dichte} where the filling 
factor $\nu$ is plotted versus $1/B$. 
At around 4\,T there is a pronounced kink in the otherwise linear
slope which is observed at the transition between the classical peak 
regime and the quantum mechanical regime in which Shubnikov-de Haas 
oscillations dominate.  Compared to the Shubnikov-de Haas 
oscillations the amplitude of oscillations superimposed upon the 
classical peak structure is strongly suppressed. We speculate that 
the weak $1/B$-periodic oscillations in the low field regime stem 
from quantization of the skipping orbits. The quantum oscillations 
are suppressed at temperatures above a few Kelvin (see lower inset 
of Fig.~\ref{dichte}) while the classical features survive up to 
temperatures of $40\,$K where they are blurred as the GaSb substrate 
becomes conducting.

%
%
 
\begin{figure}[htb] 
\begin{minipage}{8.6cm} 
\epsfxsize=8.6cm 
\epsfbox{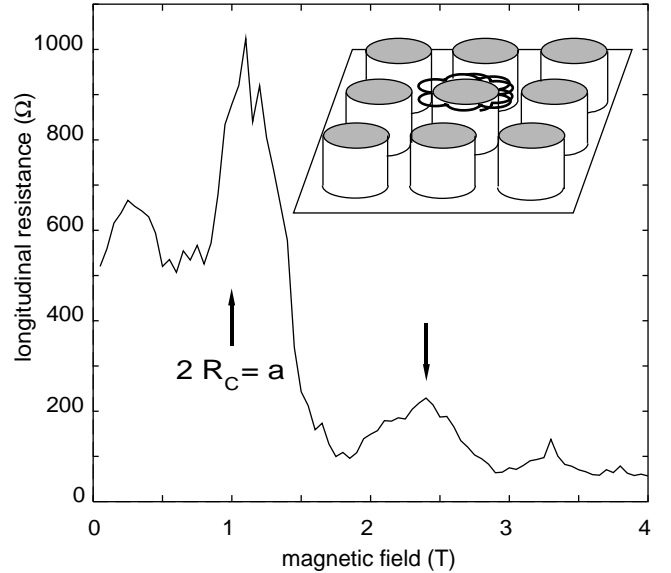} 
\end{minipage} 
\vspace{0.2cm} 
\caption{Magnetoresistance traces obtained with a simple simulation 
based on the Kubo formula. The commensurability condition $2R_c=a$ 
is shown for  $n_s=1.3 \cdot 10^{12} \,\rm{ cm}^{-2}$ used in the 
calculation. 
Inset: 
A sketch of the hard-wall potential and a rosetta orbit encircling 
an antidot. Its magnetic field position is marked with an arrow in 
the large graph.} 
\label{simu} 
\end{figure}

In order to clarify the origin of the classical maximum 
between $1.5\,$T and $3.5\,$T we carried out numerical calculations 
following recent work.\cite{Ragnar} We evaluated the Kubo formula 
\cite{Kuboformel} 
$$ 
\sigma_{ij}=\frac{e^2m^*}{\pi \hbar^2} 
\int_0^\infty \!\!\!\!{\rm d}t\, e^{-t/\tau}\langle v_i(t) v_j(0)\rangle 
$$ 
to calculate the conductivity tensor $\sigma_{ij}$ where the indices 
$i$ and $j$ stand for the $x$- and $y$-direction. The trajectories 
and the resulting velocities $v_i$ and $v_j$ were calculated 
numerically assuming a hard wall potential. The brackets denote the 
average over phase space. Previously only chaotic trajectories were 
considered in the evaluation of the  Kubo formula.\cite{Ragnar} 
Since the features we are interested in appear at relatively 
high magnetic field values, most of the phase space is occupied by 
regular non-chaotic trajectories. Therefore we included both regular 
and chaotic trajectories in our calculations.By inverting the 
conductivity tensor we obtain the experimentally determined diagonal 
resitivity $\rho_{xx}$. The results of a simulation using the simple 
model with perfectly hard walls are shown in Fig.~\ref{simu}. The 
calculated traces are remarkably close to the experimental results: 
the additional structure between 1.5 and 3.5\,T was nicely 
reproduced.\cite{fussnote} 

\begin{figure}[htb] 
\begin{minipage}{12cm} 
\epsfxsize=8.6cm 
\epsfbox{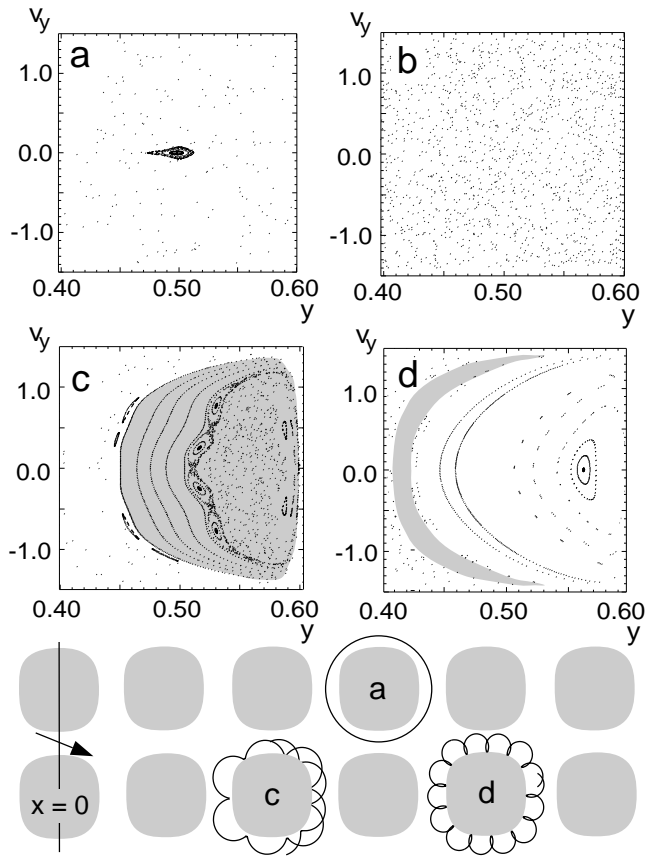} 
\end{minipage} 
\vspace{0.2cm} 
\caption{Poincar\'e surface of sections at $B$-fields labelled from  
{\sf a} to {\sf d} in Fig. \ref{uebersicht}. $y$ and $v_y$ are given 
in units of the lattice period $a$ and $\sqrt{2}\,v_F$, respectively.
Trajectories shown below correspond to shaded areas in the 
Poincar\'e surfaces of section.
Note that for the highest magnetic field ({\sf d}) the largest part 
of phase space is still occupied by stable rosetta orbits, only the 
orbits in the shaded region are unstable in an external electric 
field.
Lower left corner: 
Position of the Poincar\'e plane.} 
\label{Poincare} 
\end{figure}

To obtain information about the types of trajectories involved in 
the observed magnetoresistance anomalies we analysed the Poincar\'e 
surfaces of section at different magnetic field values. This is done 
by injecting electrons at $x = 0$ (between two antidots; see left 
hand side of the lower inset of Fig.~\ref{Poincare}) with velocity 
$v_y$ into the lattice and recording the $v_y$ and $y$ values of the 
injected electrons intersecting the line $x\,{\rm mod}\,a=0$. To 
take into account possible soft wall effects we assumed here an 
antidot potential of the form 
$$V(x,y)=V_0\left(\cos {{\pi x}\over a} \cdot \cos
{{\pi y}\over a}\right)^\beta$$ 
with $\beta = 16$. The trajectories and velocities of the injected
electrons were obtained by solving the classical equations of motion 
numerically at several magnetic fields (indicated with letters 
{\sf a} to {\sf d} in Figs. \ref{uebersicht} and \ref{Poincare}).

%
At about $1\,$T (letter {\sf a}), the commensurability condition
$2R_c = a$ is satisfied, and the Poincar\'e surface of section 
exhibits a stable island in Fig.~\ref{Poincare}a belonging to 
cyclotron orbits around one antidot. This leads to the fundamental 
commensurability peak in $\rho_{xx}$. 
The minimum in the magnetoresistance at $1.9\,$T (letter {\sf b}) is 
due to the fact that the large antidots leave no room for stable 
trajectories. This minimum is not found in antidot lattices with 
smaller AD diameters.If the magnetic field gets higher, the 
Poincar\'e surface of section~{\sf c} shows that a large fraction of 
phase space is occupied by periodic and quasi-periodic trajectories. 
These are rosetta shaped orbits, which encircle one antidot and 
remain stationary even if an electric field is applied in the
simulation, giving rise to a maximum in the magnetoresistance at 
$2.5\,$T. The situation is quite similar to pinned orbits, which are 
responsible for the commensurability peak at $2R_c=a$. The high 
field behaviour (letter {\sf d}, $B=4$\,T) of the classical 
structure is governed by the formation of unstable rosetta orbits. 
Instead of being composed more or less of semicircles, the small 
cyclotron diameter at high fields permits orbits which resemble 
circles revolving around one antidot. These orbits  tend to drift 
perpendicular to an electric field, thus reducing~$\rho_{xx}$, even 
though the stable orbits still occupy a large volume in phase space. 
 
When the magnetic field exceeds $3.5\,$T, the amplitude of the 
 SdH-oscillations increases drastically. At lower fields the 
cyclotron diameter exceeds the width of the constriction between 
adjacent antidots. Therefore, electrons can move easily from one 
edge of the sample to the other by hopping from AD to AD. As soon as 
the cyclotron diameter becomes smaller, this backscattering process 
is no longer possible and the SdH-oscillations become much more 
pronounced. From the magnetic field position of this point we 
determine this critical cyclotron diameter to be $110\,\rm{nm}$. 
If we assume a very steep antidot potential, this leads to an 
antidot diameter of $270\,\rm{nm}$ which is a 
little larger than the lithographic diameter of $250\,\rm{nm}$. 
Since on InAs surfaces the Fermi level is pinned in the conduction 
band,\cite{Leitungsband} one would not expect a depleted 
region enlarging the antidots. However, the exact pinning position 
of the Fermi level depends on the detailed conditions of the 
etch-exposed surfaces and may well be different from the bulk Fermi 
level in the 2DEG.\cite{Counterflow} 
Consequently, the enhancement of the antidot diameter can be due to 
a softer potential. 
 
In summary, we fabricated and measured antidot arrays on InAs/GaSb 
with large AD diameters. We find a broad maximum in the 
magnetoresistance at intermediate magnetic fields characterized by  
$R_c \sim b$ as lower and $2R_c \sim b$ as upper edge,
which we show to be of classical origin. The trajectories 
responsible for this maximum are identified to be rosetta orbits 
around single ADs.
We also observe weak quantum oscillations on this maximum, which we 
speculate to originate from quantized rosetta orbits. 
Finally, we estimate the effective diameter of the ADs and find that 
it is comparable with the lithographic diameter.

We would like to thank D.~Heisenberg for helpful discussions. The
 authors acknowledge the financial support of the German 
Bundesministerium f\"ur Bildung und Forschung under 
Contract \mbox{No. 01 BM 622/5.}
%
 
%

\newpage


\begin{thebibliography}{99}
\bibitem{fussnote}The calculation shows a double peak structure 
(with maxima at $\sim$2.4 and $\sim$3.3T). This double peak 
structure is not unlike the one observed in experiment, see e.g. 
the bottom trace in Fig.\ref{dichte}. While the structure can be 
ascribed to rosetta orbits we have no clear picture yet about 
the origin of the splitting.
\end{thebibliography}
\end{document}